# Analysis and calibration of star sensor's image plane displacement

Sun Yahui, Geng Yunhai, Wang Shuang

(School of Astronautics, Harbin Institute of Technology, Harbin 150001, China)

**Abstract:** Star sensor's image plane can have three kinds of displacement after a long time working in space, and the displacements are the principal point drift, incline displacement and rotation displacement. These displacements can severely decrease star sensor's measuring accuracy, therefore it's necessary to analyze and calibrate them. The previous researches have only considered the principal point drift of image plane, which is three-degree-of-freedom. In contrast, the image plane displacements under the rest three degrees of freedom, that are the incline displacement and the rotation displacement, have been modeled in this paper. These two kinds of displacement's influences on star sensor's accuracy have been analyzed. And the necessity to calibrate them has been pointed out. At last, the Extended Kalman Filter has been used to on-orbit calibrate the six-degree-of-freedom image plane displacement. And the simulation results reveal that the on-orbit calibration algorithm can effectively calibrate the image plane displacement of star sensor. The measuring accuracy of star sensor has been increased to 0.23″ after the calibration. Therefore the new six-degree-of-freedom image plane displacement model has made up the deficiency of the conventional displacement model and enhance the performance of star sensor greatly.

**CLC number:** V448.2　　**Document code:** A　　**Article ID:** 1007−2276(2014)10−3321−08

# 星敏感器像平面移位误差的分析与校正

孙亚辉，耿云海，王　爽

(哈尔滨工业大学 航天学院，黑龙江 哈尔滨 150001)

摘　要：星敏感器在长时间工作后会产生三种像平面移位误差，即主点漂移误差、倾斜误差与旋转误差。星敏感器的像平面移位误差会严重影响其测量精度。以往关于星敏感器像平面移位误差的研究仅考虑了像平面三自由度的主点漂移误差。而文中还考虑了星敏感器像平面在剩余三个自由度下的移位误差，即倾斜误差和旋转误差，从而提出了一种新的星敏感器六自由度像平面移位误差模型。最后，利用扩展卡尔曼滤波方法在轨标定了星敏感器的六自由度像平面移位误差。仿真结果显示该方法将星敏感器的测量精度大幅提高到了 0.23″，因此新的星敏感器像平面移位误差模型弥补了旧模型的不足，显著提高了星敏感器的工作性能。

**关键词：**星敏感器；　移位；　标定；　扩展卡尔曼滤波

## 0 Introduction

The aerospace area is attaching more and more importance to high-accuracy attitude determination and control of spacecraft. And the spacecraft's accuracy depends on the accuracy of its attitude determination instrument. Star sensor is the most accurate attitude determination instrument nowadays, and it's essentially an optical system working in space. A long time working in space will displace star sensor's image plane and decrease its measuring accuracy. And the micro star sensors, which are becoming more and more popular nowadays, need the consideration of image plane displacement especially.

Image plane displacement is the main factor to influence star sensor's measuring accuracy. The previous researches on image plane displacement have only considered the principal point drift[1-3], and there is less than adequate attention given to the incline displacement and the rotation displacement. Most image planes of star sensor are CCD plane[4]. And the CCD star sensor has the highest accuracy, for example, the American AST-301 star sensor's accuracy[5] is 0.18″. A high accuracy of star sensor demands consideration of as many kinds of displacement as possible.

There are many researches on the calibration method of star sensor's displacement. And the displacement can be calibrated either under the earth-based experimental environment or the space environment[6-8]. The on-orbit calibration of star sensor's displacement is meaningful inasmuch as most displacements come into being after a long time working in space. There are many algorithms to on-orbit calibrate the star sensor displacement., such as the Least Square Method (LSM), the Kalman Filter (KF), the Extended Kalman Filter (EKF), and the Unscented Kalman Filter (UKF). Considering the mathematical nature of image plane displacement, the accuracy and computation requirement of different algorithms, EKF is chosen to on-orbit calibrate the star sensor's six-degree-of-freedom displacement.

## 1 Conventional model of star sensor's image plane displacement

Star sensor can be divided into three parts, that are the optical system, the image plane and the data processing system. The conventional model of star sensor's image plane displacement can be described by Fig.1, and the principal point drift is the only kind of displacement considered in this model. In Fig.1, $OX_SY_SZ_S$ is the star sensor coordinate system without displacement, $O'X_S'Y_S'Z_S'$ is the star sensor coordinate system with displacement, $(x_0,y_0)$ is the principal point drift in $X-Y$ direction, $f_0$ is the principal point drift in $Z$ direction, $f$ is the distance between the focus point of the optical system and the image plane without displacement, $f'$ is the distance with displacement.

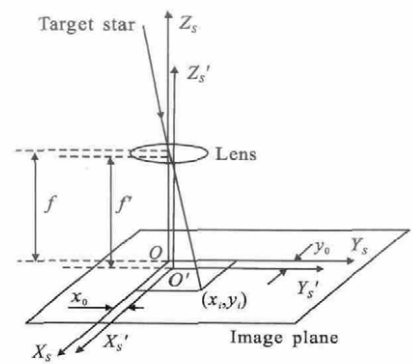

Fig.1 Conventional model of star sensor's image plane displacement

Suppose the number $i(i=1,\cdots,n)$ target star's imaging point in the star sensor coordinate system is $(x_i,y_i)$. When there is no image plane displacement, the target star's direction vector in the star sensor coordinate system is

$$W_i = \frac{1}{\sqrt{x_i^2+y_i^2+f^2}} \begin{bmatrix} -x_i \\ -y_i \\ -f \end{bmatrix} \quad (1)$$

Suppose there is image plane displacement and the image plane displacement is described by $(x_0,y_0,f_0)$, the target star's direction vector in the star sensor coordinate system is:

$$\hat{W}_i = \frac{1}{\sqrt{(x_i-x_0)^2+(y_i-y_0)^2+(f-f_0)^2}} \begin{bmatrix} -(x_i-x_0) \\ -(y_i-y_0) \\ f-f_0 \end{bmatrix} \quad (2)$$

The star sensor imagery corresponds with the ephemeris stored by the data processing system. Suppose the number $i$ target star's direction vector in the inertial coordinate system is:

$$\hat{V}_i = \begin{bmatrix} \cos\alpha_i \cos\delta_i \\ \sin\alpha_i \cos\delta_i \\ \sin\delta_i \end{bmatrix} \quad (3)$$

in which $\alpha_i$ and $\delta_i$ are respectively the target star's right ascension and declination. Based on the principal of angular distance's equality, the relationship between the target star's direction vector in the star sensor coordinate system and the corresponding one in the inertial coordinate system can be illuminated as

$$\hat{W}_i^T \hat{W}_j = \hat{V}_i^T \hat{V}_j \quad (4)$$

Then with the principal of angular distance's equality, the attitude of satellite can be determined by matching the stellar map.

Fig.2 and Fig.3 are the analysis results of star sensor image plane's principal point drift, in which $r=\sqrt{x^2+y^2}$ is the distance between the imaging point and the principal point, $dr$ is the principal point drift in $X-Y$ direction, $df$ is the principal point drift in $Z$ direction. It can be easily seen from Fig.2 and Fig.3 that the measuring error caused by the principal point drift in $X-Y$ direction is becoming bigger as the imaging point access the principal point, while the measuring error caused by the principal point drift in $Z$ direction is becoming smaller as the imaging point access the principal point. When $r=15$ mm, both a 0.000 3 mm drift in $X-Y$ direction and a 0.001 mm drift in $Z$ direction can engender a measuring error of star sensor bigger than 1 second of arc. Hence the image plane's principal point drift has a big influence on star sensor's measuring accuracy, and the principal point drift must be calculated.

However, the principal point drift is only one part of star sensor's image plane displacement. The other two kinds of displacement are the incline displacement and the rotation displacement, and they also have an un-negligible influence on star sensor's measuring accuracy. The six-degree-of-freedom image plane displacement includes the principal point drift, the incline displacement and the rotation displacement. With the enhancement of star sensor's measuring accuracy and the emergence of micro star sensor, the six-degree-of-freedom image plane displacement should be considered and calibrated, not only the principal point drift.

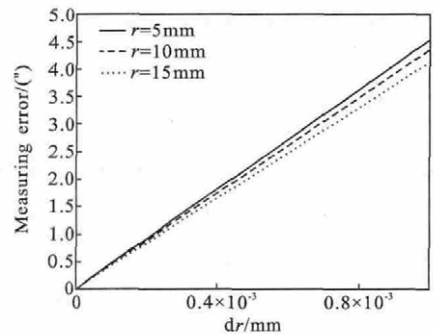

Fig.2 Analysis of image plane's principal point drift in $X-Y$ direction

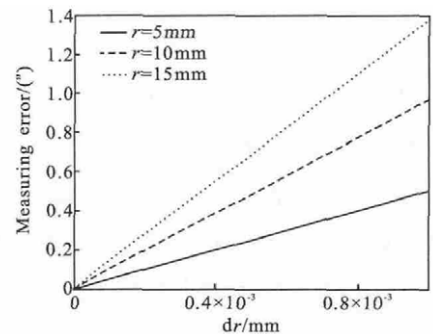

Fig.3 Analysis of image plane's principal point drift in $Z$ direction

## 2　Incline displacement of star sensor's image plane

Satellite may endure occasional oscillations when it is working in space[9], and the irregular thermal environment in space can also displace star sensor's image plane[10-11]. Many different factors cause the incline displacement of star sensor's image plane.

Fig.4 is the model of image plane's incline displacement, in which $S_a$ is the image plane without incline displacement, $S_b$ is the image plane with incline displacement. $OX_aY_aZ_a$ is the star sensor coordinate system without incline displacement, $OX_bY_bZ_b$ is the star sensor coordinate system with incline displacement, and the axis $Z_a$ is the optical axis of the optical system.

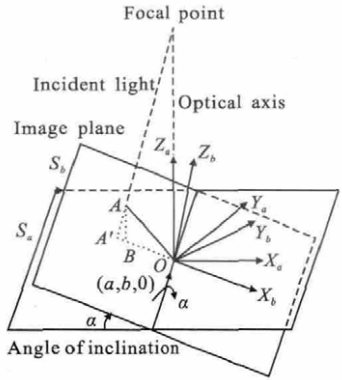

Fig.4 Incline displacement model of star sensor's image plane

Suppose the inclination vector is $(a,b,0)$ in the coordinate system $OX_aY_aZ_a$, and the angle of inclination is $\alpha$. As a result, the direction cosine matrix of the coordinate system $OX_bY_bZ_b$ relative to the coordinate system $OX_aY_aZ_a$ is

$$C_{ba}=\begin{bmatrix} \cos\alpha+a^2(1-\cos\alpha) & ab(1-\cos\alpha) & -b\sin\alpha \\ ab(1-\cos\alpha) & \cos\alpha+b^2(1-\cos\alpha) & a\sin\alpha \\ b\sin\alpha & -a\sin\alpha & \cos\alpha \end{bmatrix} \quad (5)$$

And $C_{ba}^{\mathrm{T}}=C_{ba}^{-1}$. Let $(l,m,n)$ be the coordinate of imaging point $A$ in the coordinate system $OX_aY_aZ_a$, $(x,y,0)$ be the coordinate of imaging point $A$ in the coordinate system $OX_bY_bZ_b$, and there will be equations below.

$$\begin{bmatrix} x \\ y \\ 0 \end{bmatrix} = C_{ba}\begin{bmatrix} l \\ m \\ n \end{bmatrix} \quad (6)$$

$$\begin{bmatrix} l \\ m \\ n \end{bmatrix} = C_{ba}^{\mathrm{T}}\begin{bmatrix} x \\ y \\ 0 \end{bmatrix} \quad (7)$$

The coordinate of imaging point $A$ in the coordinate system can be gotten with equation 7.

$$\begin{bmatrix} l \\ m \\ n \end{bmatrix} = \begin{pmatrix} [\cos\alpha+a^2(1-\cos\alpha)]x+ab(1-\cos\alpha)y \\ ab(1-\cos\alpha)x+[\cos\alpha+b^2(1-\cos\alpha)]y \\ -b\sin\alpha x+a\sin\alpha y \end{pmatrix} \quad (8)$$

The point $A$ in Fig.4 is the imaging point with the incline displacement, and the imaging point will change to $A'$ if there is no incline displacement. With the point $A'$'s coordinate $(l,m,n)$ in the coordinate system $OX_aY_aZ_a$, the point $A'$'s coordinate $(x',y',0)$ in the coordinate system $OX_aY_aZ_a$ can be gotten, and $(x',y',0)$ is the calibrated imaging point.

$$(x',y',0)=\left(\frac{f}{f-n}l,\frac{f}{f-n}m,0\right) \quad (9)$$

Suppose the inclination vector $(a,b,0)$ is an unit vector, and $b=\sqrt{1-a^2}$. Then the parameters of image plane's incline displacement are $(a,\alpha)$. It can be seen from the parameters that the incline displacement of star sensor's image plane is a two-degree-of-freedom displacement.

Define the angle of incidence as the angle between target star's direction vector and the optical axis, and the angle of incidence in Fig.5 is $17.45°$. It can be seen from Fig.5 that when the angle of inclination $\alpha=0.02°$, the measuring error of star sensor is above $15''$. And every kind of displacement which causes a measuring error bigger than $0.5''$ should be considered and calibrated. Therefore the incline displacement of image plane also has a big influence on the measuring accuracy of star sensor. The incline displacement of star sensor's image plane should be calibrated.

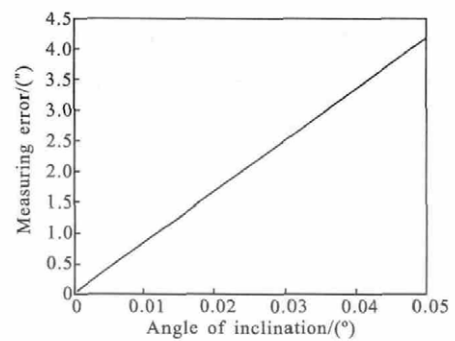

Fig.5 Analysis of image plane's incline displacement

## 3 Rotation displacement of star sensor's image plane

The terrible space environment can causes rotation displacement of star sensor's image plane. And Fig.6 is the model of star sensor image plane's rotation displacement.

In Fig.6, $OX_aY_aZ$ is the coordinate system of star sensor without rotation displacement, $OX_bY_bZ$ is the coordinate system of star sensor with rotation displacement. If the rotation vector is $Z$ axis and the angle of rotation is $\psi$, the direction cosine matrix of the coordinate system $OX_bY_bZ$ relative to the coordinate system $OX_aY_aZ$ is:

$$C_Z(\psi)=\begin{bmatrix} \cos\psi & \sin\psi & 0 \\ -\sin\psi & \cos\psi & 0 \\ 0 & 0 & 1 \end{bmatrix} \quad (10)$$

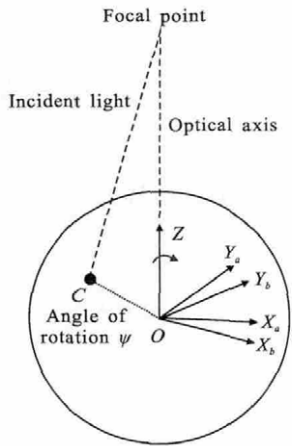

Fig.6 Rotation displacement model of star sensor's image plane

The coordinate of imaging point $C$ is $(x',y',0)$ in the coordinate system $OX_aY_aZ$, and the coordinate of imaging point $C$ is $(x',y',0)$ in the coordinate system $OX_bY_bZ$. Then the calibrated imaging point is:

$$\begin{bmatrix} x' \\ y' \\ 0 \end{bmatrix} = C_Z^T(\psi)\begin{bmatrix} x \\ y \\ 0 \end{bmatrix} = \begin{pmatrix} \cos\psi x - \sin\psi y \\ \sin\psi x + \cos\psi y \\ 0 \end{pmatrix} \quad (11)$$

It's easy to see from the equation above that the rotation displacement of star sensor's image plane is an one-degree-of-freedom displacement, and the displacement parameter is $\psi$.

Fig.7 is the analysis results of image plane's rotation displacement. The angle of incidence in Fig.7 is $17.45°$. It can be seen from Fig.7 that the rotation displacement's influence on star sensor's measuring accuracy is also significant. It is necessary to calibrate the star sensor image plane's rotation displacement.

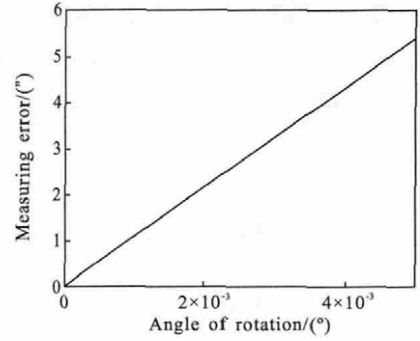

Fig.7 Analysis of image plane's rotation displacement

## 4　Sin-degree-of-free DOM displacement of star sensor's image plane

The six-degree-of-freedom image plane displacement includes the three-degree-of-freedom principal point drift, the two-degree-of-freedom incline displacement and the one-degree-of-freedom rotation displacement. Three sub-kinds of displacement should all be calibrated to increase the performance of star sensor.

Fig.8 is the six-degree-of-freedom displacement model of star sensor's image plane, in which the point $A'$ is the final calibrated imaging point. $OX_aY_aZ_a$

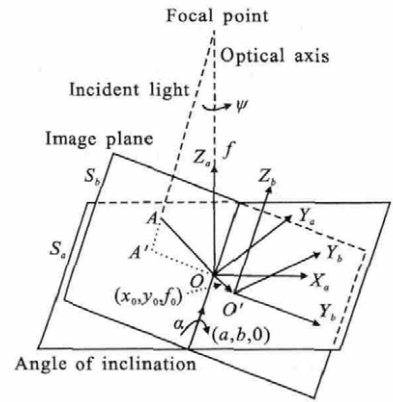

Fig.8 Six-degree-of-freedom displacement model of star sensor's image plane

is the coordinate system without image plane displacement, $O'X_aY_aZ_a$ is the coordinate system with image plane displacement. Equation (12) and (13) are the calibration equations of star sensor's six-degree-of-freedom image plane displacement, in which $(x,y,0)'$ is the coordinate of the real imaging point $A$ in $O'X_bY_bZ_b$, $(x',y',f')$ is the coordinate of the real imaging point $A$ in $OX_aY_aZ_a$, $(x_{new},y_{new},0)'$ is the coordinate of the calibrated imaging point $A'$ in $OX_aY_aZ_a$, $(0,0,f)'$ is the coordinate of the calibrated focus point in $OX_aY_aZ_a$.

$$\begin{bmatrix} x' \\ y' \\ f' \end{bmatrix} = C_{ba}^T C_Z^T(\psi) \begin{bmatrix} x \\ y \\ 0 \end{bmatrix} + \begin{bmatrix} x_0 \\ y_0 \\ f_0 \end{bmatrix} \quad (12)$$

$$\begin{cases} x_{new} = x' \cdot \dfrac{f}{f-f'} \\ y_{new} = y' \cdot \dfrac{f}{f-f'} \end{cases} \quad (13)$$

## 5 On-orbit calibration of star sensor's image plane displacement

On-orbit calibration of star sensor's image plane displacement can effectively increase star sensor's measuring accuracy. There are many kinds of calibration algorithm to calibrate star sensor displacements. Considering the mathematical nature of image plane displacement, the accuracy and computation requirement of different algorithms, EKF is chosen to on-orbit calibrate star sensor's six-degree-of-freedom image plane displacement. EKF can efficiently filter the nonlinear system under the white Gaussian noise.

The theoretical basis of star sensor's on-orbit calibration is the principle of angular distances' equality. And the calibration target is $(a,\alpha,\psi,x_0,y_0,f_0)$. Based on the calibration equations in Section 5, the system functions of on-orbit calibration are:

$$\begin{cases} \xi(k+1)=f(k,\xi(k))+w(k) \\ y(k)=h(k,\xi(k))+v(k) \end{cases} \quad (14)$$

In which $\xi(k)$ is the difference between the real displacement parameters and the calibrated displacement parameters, $y(k)$ is the difference between the real angular distances and the calibrated angular distances after $k$ times of iteration. They can be evinced as follows

$$\xi(k)=[\Delta a\ \Delta\alpha\ \Delta\psi\ \Delta x_0\ \Delta y_0\ \Delta f_0]' \quad (15)$$

$$y(k)=\begin{bmatrix} W_1^T W_2 - V_1^T V_2 \\ \vdots \\ W_1^T W_{num} - V_1^T V_{num} \\ W_2^T W_3 - V_2^T V_3 \\ \vdots \\ W_{num-1}^T W_{num} - V_{num-1}^T V_{num} \end{bmatrix} \quad (16)$$

where is the number of imaged target stars used in this calibration. In the system functions, $w(k)$ is the systematic noise and $v(k)$ is the measured noise. They conform to the following rules.

$$\begin{cases} E[w(k)]=0,\ E[w(k)w(k)^T]=Q^w(k) \\ E[v(k)]=0,\ E[v(k)V(k)^T]=Q^v(k) \\ E[w(k)v(k)^T]=0, \end{cases} \quad (17)$$

The state variable can be written as

$$\hat{\xi}(k+1)=f(k,\hat{\xi}(k))+N(k)[y(k)-h(k,\hat{\xi}(k))] \quad (18)$$

And the processes of measurement update and time update are:

$$\begin{aligned} N(k)&=F(k,\hat{\xi}(k))P(k)H^T(k,\hat{\xi}(k))\times \\ &\quad [H(k,\hat{\xi}(k))P(k)H^T(k,\hat{\xi}(k))+Q^v(k)]^{-1}, \\ P(k+1)&=F(k,\hat{\xi}(k))P(k)F^T(k,\hat{\xi}(k))+Q^w(k)-N(k) \\ &\quad [Q^v(k)+H(k,\hat{\xi}(k))P(k)H^T(k,\hat{\xi}(k))]N^T(k) \end{aligned} \quad (19)$$

Where $F(k,\hat{\xi})$ and $H(k,\hat{\xi})$ are respectively the Jacobi matrix of $f(k,\hat{\xi})$ and $h(k,\hat{\xi})$.

$$F(k,\hat{\xi})=\frac{\partial}{\partial\xi}f(k,\hat{\xi})|_{\xi=\hat{\xi}}=I \quad (20)$$

$$H(k,\hat{\xi})=\frac{\partial}{\partial\xi}h(k,\hat{\xi})|_{\xi=\hat{\xi}}= \begin{bmatrix} \frac{\partial}{\partial a}W_1^T W_2 & \cdots & \frac{\partial}{\partial f_0}W_1^T W_2 \\ \vdots & \cdots & \vdots \\ \frac{\partial}{\partial a}W_{num-1}^T W_{num} & \cdots & \frac{\partial}{\partial f_0}W_{num-1}^T W_{num} \end{bmatrix} \quad (21)$$

Fig.9 is the EKF calibration process of star sensor's six-degree-of-freedom image plane displacement. The displacement parameters can be calibrated with the initial value $P(0)$ and $\hat{\xi}(0)$.

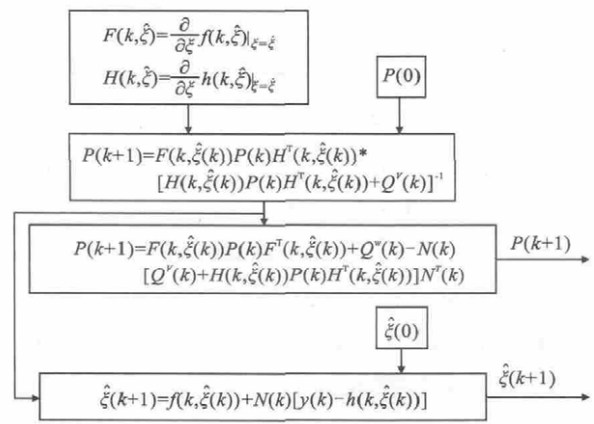

Fig.9 Flow chart of extended Kalman filter

## 6　Simulation and analysis

In the simulation of on-orbit calibration, the focal length of star sensor is 45 mm. And the six displacement parameters are fixed as follows.

$$(a,\alpha,\psi,x_0,y_0,f_0)=(0.5,0.02°,0.01°,0.02\text{ mm},$$
$$0.02\text{ mm},0.02\text{ mm}) \quad (22)$$

20 groups of target stars' imaging data are used in the simulation, and a certain amount of white Gaussian noise is mixed with the imaging data. The simulation results of on-orbit calibration are showed by Fig.10 and Fig.11.

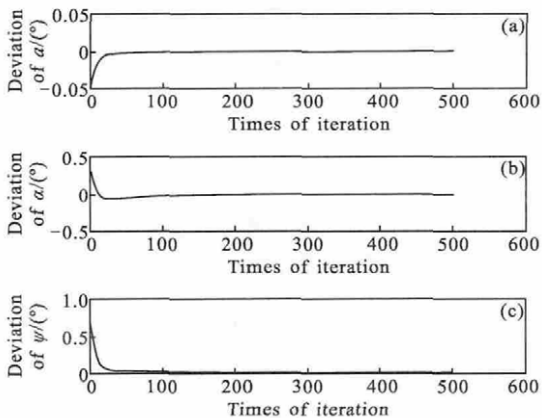

Fig.10　Calibration result of displacement parameters (a)

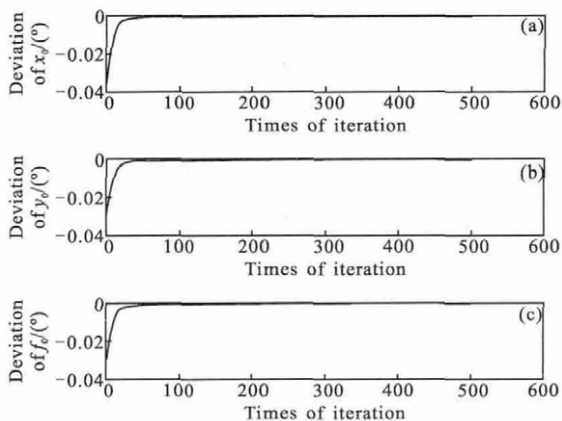

Fig.11　Calibration result of displacement parameters (b)

It can be easily seen from Fig.10 and Fig.11 that EKF is able to calibrate star sensor's six-degree-of-freedom image plane displacement effectively. And all the six calibration results become steady when the time of iteration is above 100. The ultimate on-orbit calibration result is

$$[\Delta a\ \Delta\alpha\ \Delta\psi\ \Delta x_0\ \Delta y_0\ \Delta f_0]=(0.000\,2,-0.000\,1'',0.000\,12'',$$
$$-0.000\,11\text{ mm},-0.000\,12\text{ mm},-0.000\,12\text{ mm}) \quad (23)$$

The measuring accuracy of star sensor has been increased to $0.23''$ after the calibration. Therefore the model of six-degree-of-freedom image plane displacement can perfectly describe the image plane displacement of star sensor, and the calibration method above can effectively calibrate it.

## 7　Conclusion

The conventional model of star sensor's image plane displacement has been analyzed and its deficiency has been pointed out. The models of image plane's incline displacement and rotation displacement have been built and their influences on star sensor's measuring accuracy have been analyzed. The model of star sensor's six-degree-of-freedom image plane displacement has been built ultimately and the Extended Kalman Filter is utilized to calibrate the displacement. The simulation results show that the calibration method can effectively calibrate star sensor's image plane displacement and increase star sensor's measuring accuracy. Therefore the six-degree-of-freedom model and the on-orbit calibration method are practically meaningful. Future work researching on star sensor's working environment is needed to determine the magnitude of image plane displacement and further certify the six-degree-of-freedom model of image plane displacement.

**References：**

[1]　Singla P, Griffith D T, Crassidis J L, et al. Attitude determination and autonomous on-orbit calibration of star tracker for the gifts mission [J]. ***Advances in the Astronautical Sciences***, 2002, 112: 19−38.

[2]　Zhong H, Yang M, Lu X. Calibration method of star sensor [J]. ***Acta Optica Sinica***, 2010, 30(5): 1343−1348.

[3]　He P, Liang B, Zhang T, et al. Calibration method for wide field of view star sensor [J]. ***Acta Optica Sinica***, 2011, 31 (10): 1023001.

[4]　Liu L, Zhang L, Zheng X, et al. Current situation and

development trends of star sensor technology [J]. *Infrared and Laser Engineering*, 2007, 36: 529−533.

[5] Van B R, Swanson D, Boyle P. Flight performance of the spitzer space telescope AST−301 autonomous star tracker[C]// 28th Annual AAS Rocky Mountain Guidance and Control Conference, 2005.

[6] Yuan Y, Geng Y, Chen X. Autonomous on-orbit calibration algorithm of star sensors with gyros[J]. *Systems Engineering and Electronics*, 2008, 30(1): 120−123.

[7] Shen J, Zhang G, Wei X. On-orbit calibration of star sensor based on kalman filter [J]. *Acta Aeronautica ET Astronautica Sinica*, 2010, 31(6): 1220−1224.

[8] Ahmed M T, Farag A A. Differential methods for nonmetric calibration of camera lens distortion [J]. *Computer Vision and Pattern Recognition*, 2001, 2: II−477−II−482 vol. 2.

[9] Wang J, Jiao Y, Zhou H, et al. Star sensor attitude measuring data processing technique in condition of complex satellite dithering[J]. *Journal of Electronics and Information Technology*, 2010, 32(8): 1885−1891.

[10] Applewhite R W, Telkamp A R. The effects of thermal gradients on the Mars Observer Camera primary mirror [J]. *Aerospace Sensing. International Society for Optics and Photonics*, 1992, 376−386.

[11] Griffith D T, Singla P, Junkins J L. Autonomous on−orbit calibration approaches for star tracker cameras [J]. *Advances in the Astronautical Sciences*, 2002, 112: 39−57.